# Refinement of Nb$_3$Sn grain size by the generation of ZrO$_2$ precipitates in Nb$_3$Sn wires

X. Xu,[1,a)] M. Sumption,[1] X. Peng,[2] and E. W. Collings[1]

[1]*Department of Materials Science and Engineering, the Ohio State University, Columbus, OH43210, USA*

[2]*Hyper Tech Research Incorporated, 539 Industrial Mile Road, Columbus, OH 43228, U.S.A*

In this letter we demonstrate that if oxygen can be properly supplied to (Nb-Zr)-Sn wires, ZrO$_2$ precipitates will form during the heat treatment, refining the Nb$_3$Sn grain size markedly. Here, a Nb$_3$Sn subelement was fabricated in which Nb-1Zr alloy was used, and oxygen was supplied via SnO$_2$ powder. The results showed that such a design could supply sufficient oxygen to internally oxidize the Zr in the Nb-1Zr alloy, and that the sample reacted at 650 °C had grain sizes of ~45 nm, less than half the size of the grains in present Nb$_3$Sn conductors. Magnetic measurements showed that the peak of the pinning force vs. field ($F_p$-$B$) curve was shifted to ~0.3$B_{irr}$ (the irreversibility field).

Nb$_3$Sn superconductors are of significant interest for applications in particle accelerators, undulators, fusion reactor designs, nuclear magnetic resonance (NMR) machines, and research magnets.[1] The planned upgrade of the interaction region magnets of the Large Hadron Collider (LHC) to meet the goal of higher luminosity requires Nb$_3$Sn conductors of high performance, especially high critical current density $J_c$ at up to 15 T.[2]

Improvement of the high field $J_c$ of Nb$_3$Sn conductors can be realized in two ways: increasing the irreversibility field ($B_{irr}$) and enhancing the pinning capacity. The potential for the improvement in the $B_{irr}$ of Nb$_3$Sn conductors is limited, as the $B_{irr}$s of the state-of-the-art Nb$_3$Sn strands at 4.2 K are normally 24-26 T,[3-5] which is quite close to the limit of the upper critical field ($B_{c2}$) of ternary Nb$_3$Sn, ~27 T at 4.2 K (which corresponds to ~31 T at 0 K).[6] The greatest potential for Nb$_3$Sn conductor development, on the other hand, lies in improving the pinning capacity. It has long been recognized that grain boundaries are the main flux line pinning centers in Nb$_3$Sn. Present Nb$_3$Sn conductors, with grain sizes of typically 100-200 nm, follow flux shear behavior, illustrated by the fact that the bulk pinning force vs. field ($F_p$-$B$) curve peaks at 0.2$B_{irr}$.[7] Dietderich's experiment[8] on Nb$_3$Sn films fabricated by electron-beam co-evaporation showed that if the Nb$_3$Sn grain size is refined to 15-30 nm, the peak of the $F_p$-$B$ curve could be shifted from 0.2$B_{irr}$ to 0.5$B_{irr}$, which can improve the 12 T $J_c$ by a factor of three.[8]

---

a) Author to whom correspondence should be addressed. Electronic mail: xu.452@osu.edu.

The primary method that has been used to control the Nb$_3$Sn grain size has been lowering the reaction temperature. However, further lowering of the reaction temperature relative to the current level (625-650 °C) is inadvisable as it drives the Nb$_3$Sn phase off-stoichiometry and also requires excessive reaction time. On the other hand, the internal oxidation method that was successfully used in Nb$_3$Sn films[9] offers an alternative. The film approach used a Nb-Zr foil into which oxygen was supplied by anodization and subsequent annealing; after being coated with Cu-Sn, the foil was reacted at 1050 °C to form Nb$_3$Sn with ZrO$_2$ precipitates. It was reported that with sufficient oxygen content, the Nb$_3$Sn grain size could be refined by an order of magnitude.[9]

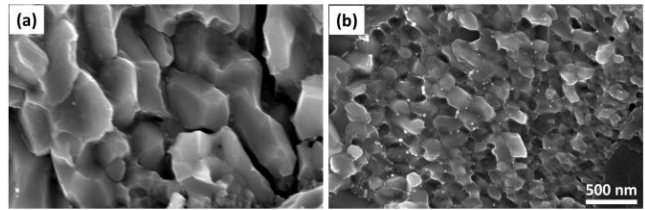

Fig. 1. Fracture SEM images of type A samples reacted at 850 °C for 10 min in (a) pure argon and (b) argon-oxygen atmospheres.

Transferring this method to Nb$_3$Sn wires, however, has not yet been realized, mainly because pre-dissolving oxygen in Nb-Zr markedly increases its strength and decreases its ductility, making the wires hard to process. To circumvent the usage of the Nb-O alloy, Zeitlin added SnO$_2$ powder into the Sn core of a mono-element internal-tin strand, expecting the Nb-Zr to reduce SnO$_2$ during heat treatment and take up the oxygen,[10] because according to the Ellingham diagram, Nb and Zr have much stronger affinity to oxygen than does Sn. However, no grain refinement was observed in his samples reacted at or below 850 °C.

To find out whether this was due to a defect in the experimental design or alternatively because generating ZrO$_2$ precipitates in Nb$_3$Sn wires is inherently impossible at low reaction temperatures, we fabricated a series of composites, starting with a Nb-1Zr alloy tube surrounding a Cu/Sn core, here denoted composite A. With such a design the Nb-1Zr alloy was exposed to the atmosphere during heat treatment, so oxygen could be supplied to the Nb-1Zr externally. For comparison various samples were given the same heat treatments in pure argon and in argon-oxygen atmospheres at various temperatures from 650 to 850 °C.

Fracture scanning electron microscopy (SEM) images of the samples reacted at 850 °C without and with oxygen are shown in Fig. 1. It is clear that the grain size of the sample reacted in argon-oxygen atmosphere is substantially smaller. From the transmission electron microscopy (TEM) image shown in Fig. 2, we can clearly see intra-granular ZrO$_2$ particles of various sizes.[9] Fig. 3 presents the dependence of Nb$_3$Sn grain size for type A samples on reaction temperature. The grain sizes were calculated by the method described in Ref. [11]. At all of the investigated temperatures the grain sizes of the samples reacted in argon-oxygen mixtures are much smaller than those reacted in the absence of oxygen. Similar to standard Nb$_3$Sn strands, ZrO$_2$ stabilized Nb$_3$Sn grains become smaller as the reaction temperature is reduced. We also noticed that for the samples containing ZrO$_2$ there is a grain size gradient across the Nb$_3$Sn layer. For example, the grain size of the sample reacted at 650 °C changes from ~55 nm near the Cu-Sn core to ~40 nm near the unreacted Nb.

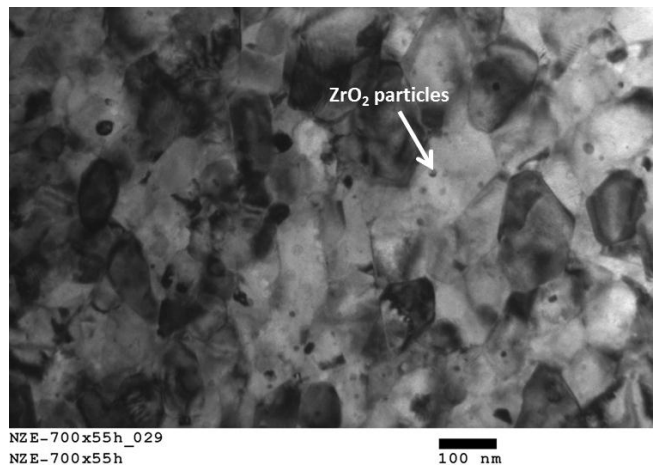

Fig. 2. TEM image of sample A reacted at 700 °C for 55 hours in an argon-oxygen atmosphere. An example of a ZrO$_2$ particle is marked.

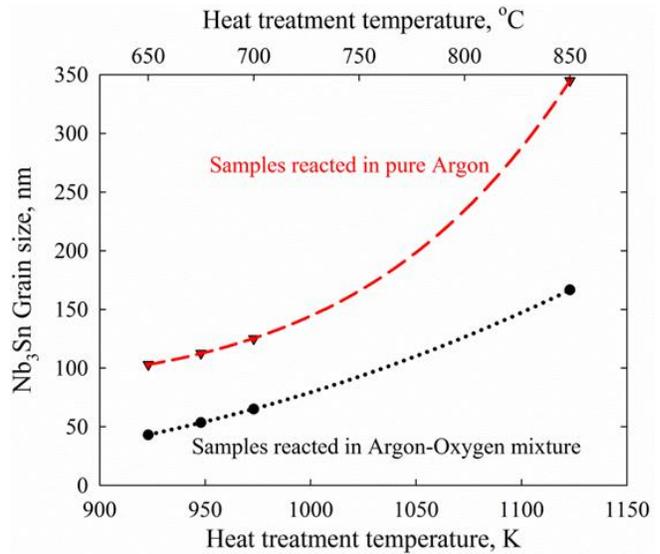

Fig. 3. Average Nb$_3$Sn grain size as a function of reaction temperature for type A samples reacted in pure argon and in argon-oxygen atmospheres. The dashed lines are exponential fits to the data.

Since grain size variations across the Nb$_3$Sn layers in standard Nb$_3$Sn wires are typically small (excluding cases where grain morphology changes),[12] we think this grain size gradient could be due to an oxygen concentration gradient across the Nb-1Zr layer prior to Nb$_3$Sn formation. One possible mechanism of the grain refinement due to the formation of ZrO$_2$ particles was proposed by Rumaner *et al.*[9]: Zr and O dissolve readily into Nb, but have limited solubility in Nb$_3$Sn, so during the Nb$_3$Sn layer growth they are expelled and precipitated out as fine ZrO$_2$ particles. The ZrO$_2$ precipitates are coherent with the host Nb$_3$Sn lattice and can inhibit the Nb$_3$Sn grain boundaries from migrating.

The above study demonstrates that ZrO$_2$ precipitates can form and the grain size can be refined in (Nb-Zr)$_3$Sn strands provided that sufficient oxygen is absorbed by Nb-Zr before it is transformed to Nb$_3$Sn. Returning to Zeitlin's study, we speculate that the reason why no grain refinement occurred was that the intact Cu layer that surrounded the Sn+SnO$_2$ core in his conductor[10] had, according to the Ellingham diagram, a lower affinity to oxygen than did Sn, thus prevented SnO$_2$ from being reduced. In other words, the "inert" Cu layer blocked the path of oxygen transfer. It should then be possible to inject oxygen from within the composite, as long as the SnO$_2$ powder reaches the Nb-Zr directly, without the intervening copper (itself needed for A15 formation). To verify the feasibility of this approach, we fabricated a second composite, denoted B, by filling SnO$_2$ powder into a Cu encased Nb tube and drawing it down to 0.7 mm diameter, and subsequently reacted several segments at various temperatures. The oxygen content in the Nb was determined by measurement of the critical temperature ($T_c$): according to previous work,[13, 14] the $T_c$ of Nb drops by about 0.93 K for each 1 at.% O dissolved in Nb. The magnetic moment vs. temperature ($m$-$T$) curves for the samples after various heat treatments are shown in Fig. 4(a). It can be seen that the oxygen content increases with reaction temperature – which could be because of either increased absorption rate or increased oxygen solubility in Nb with temperature, reaching about 2 at.% for the sample reacted at 500 °C. It is also interesting to note that for the samples reacted at and above 550 °C, $NbO_2$ and NbO compounds formed after a certain reaction time, an example shown in Fig. 4 (b). We speculate that due to excessive $SnO_2$ amount in this composite B, as the oxygen content in the Nb saturated, Nb-O compounds began to form. Thus, this study warns against using excessive $SnO_2$ for oxygen supply.

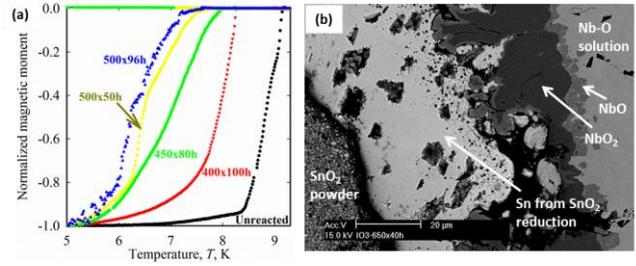

Fig. 4. (a) $M$-$T$ curves of samples B reacted at various temperatures, and (b) SEM image of sample B reacted at 650 °C for 40 h, showing the Nb-O compounds.

With the knowledge that it is feasible to use $SnO_2$ powder as the oxygen source, we then fabricated a third composite, denoted C, along this line, by filling a Nb-1Zr tube with a Sn/Cu/$SnO_2$ tubular layer structure (that is, the $SnO_2$ powder is positioned between the Cu/Sn core and the Nb-1Zr tube wall), as shown in Fig. 5. For comparison purposes, we also fabricated an analog, D, with $NbO_2$ powder instead of $SnO_2$, and both were reacted at 650 °C for 150 h. The fracture SEM images of the reacted samples are shown in Fig. 6. The average grain sizes of samples C (with $SnO_2$) and D (with $NbO_2$) are 43 nm and 91 nm, respectively. In sample C there are a large fraction of grains with size smaller than 30 nm. The magnetic moment vs. temperature ($m$-$T$) curves in Fig. 7 (a) shows that the $T_c$ of the unreacted Nb-1Zr in sample D is 9.1 K, whereas the value for sample C is 6.6 K, indicating that the $NbO_2$ powder failed to supply much oxygen, while in sample C nearly 3 at.% oxygen was absorbed by the Nb-1Zr alloy. Magnetic moment versus field measurements at 4.2 K (in the perpendicular direction) were performed on the reacted samples, and magnetic $Nb_3Sn$ layer $J_c$s were calculated. The $F_p$-$B$ curves for both samples are shown in Fig. 7 (b). The curves were fitted to the universal scaling law: $F_p=Kb^p(1-b)^q$, where $K$ is a pre-factor related to the maximum pinning force $F_{p,max}$, and $b=B/B_{irr}$. The fitted $B_{irr}$ values of samples C and D are respectively 20.5 and 21.3 T, relatively low because the samples were based on binary $Nb_3Sn$ and were under-reacted.[15] Fig. 7 (c) presents the $F_p/F_{p,max}$ vs. $B/B_{irr}$ curves of samples C and D, peaking at ~0.3$B_{irr}$ and ~0.2$B_{irr}$, respectively. Sample D more or less follows the typical flux shear behavior; for sample C, on the other hand, an apparent shift of the $F_p$-$B$ curve to higher field is observed. The high fraction of grains with size of sub-30 nm in sample C perhaps accounts for the shift of the $F_p$-$B$ curve. It is interesting to note from Fig. 7 (b) that although sample C has smaller grain size, its $F_{p,max}$ is smaller than that of sample D. Similar phenomenon is also seen in the data of Dietderich's thin films.[8] Possibly this is because sample D has higher upper critical field $B_{c2}$, and $F_{p,max}$ is proportional to $B_{c2}^{2.5}$.[7] Another potential reason is that the pinning function $F_p$ changes its form as the pinning mechanism switches, as explained by Dew-Hughes' theory.[16] The 12 T layer $J_c$ of sample C is around 6.1 kA/mm$^2$. However, a fully reacted Ti doped strands can have a $B_{irr}$ of ~25 T.[5] With such a $B_{irr}$, we estimate that the 12 T layer $J_c$ might reach ~10kA/mm$^2$. It is worth noting that a secondary benefit that the shift of the $F_p$-$B$ curve to higher field brings is a lower $J_c$ at low field, which benefits the low field stability.[17]

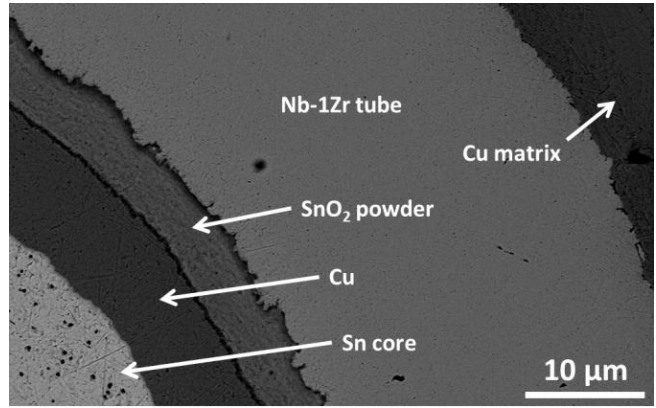

Fig. 5. SEM image of composite C. The black dots in the Sn core are due to polishing.

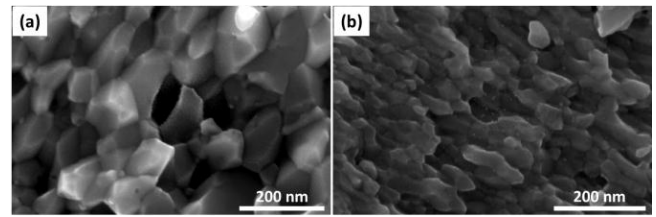

Fig. 6. Fracture SEM images of (a) sample D (with $NbO_2$) and (b) sample C (with $SnO_2$) reacted at 650 °C for 150 h.

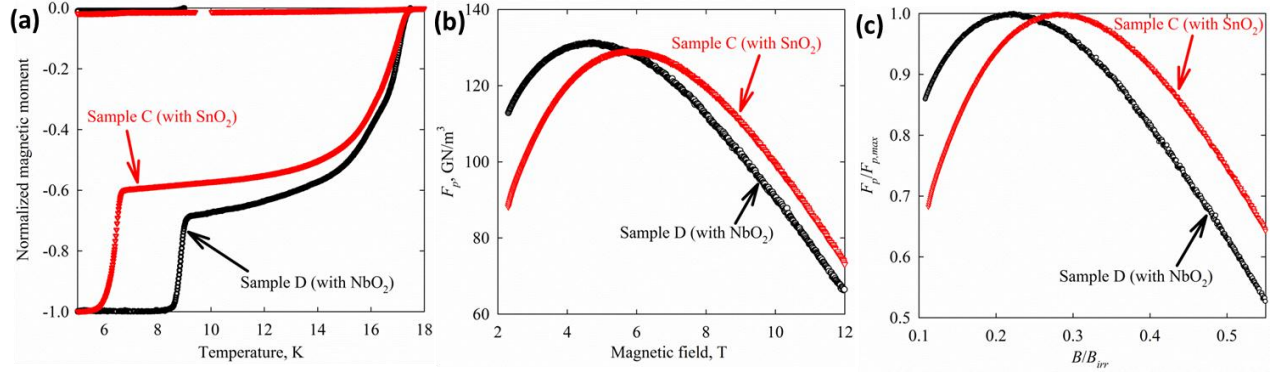

Fig. 7. The (a) $m$-$T$, (b) $F_p$-$B$, and (c) reduced $F_p$-$B$ curves of samples D and C reacted at 650 °C for 150 h.

In summary, this work explores the way to apply the internal oxidation method to $Nb_3Sn$ wires for grain size refinement. It was shown that $SnO_2$ powder can be used as oxygen source, but must be positioned in a way that no Cu blocks the path of oxygen to Nb-Zr alloy. This led to the fabrication of composite C, which, when reacted at 650 °C, achieved an average grain size of ~43 nm as well as a shift of the $F_{p,max}$ to ~0.3$B_{irr}$. It is possible that lowering the reaction temperature to 625 °C would lead to smaller grain size with the peak of $F_p$-$B$ curve shifting to even higher fields. Beyond this, it is also of interest to use higher Zr contents to see if grain sizes can be reduced even further and to repeat this work with samples with Ti additions via Sn-Ti alloy for $B_{irr}$ optimization. In light of the results obtained, we anticipate that this approach could lead to substantial improvement in the performance of $Nb_3Sn$ conductors at both high and low fields.

This work was funded by the US Department of Energy, Division of High Energy Physics, Grant No. DE-FG02-95ER40900, and a DOE Contract Numbers DE-SC0010312.